# Title: High-quality imaging of large areas through path-difference ptychography

**Authors:** Jizhe Cui[1,2,3†], Yi Zheng[1,2,3†], Kang Sun[1,2,3†], Wenfeng Yang[1,2,3], Haozhi Sha[1,2,3] and Rong Yu[1,2,3*]

Affiliations:

[1]School of Materials Science and Engineering, Tsinghua University, Beijing 100084, China.

[2]Key Laboratory of Advanced Materials of Ministry of Education, Tsinghua University, Beijing 100084, China.

[3]State Key Laboratory of New Ceramics and Fine Processing, Tsinghua University, Beijing 100084, China.

*Corresponding author. Email: ryu@tsinghua.edu.cn

†These authors contributed equally to this work.

**Abstract:**

Tilting planar samples for multi-zone-axes observation is a routine procedure in electron microscopy. However, this process invariably introduces optical path differences in the electron beam across different sample positions, significantly compromising image quality, particularly over large fields of view. To address this challenge, we developed path difference ptychography (PDP), a method capable of decoupling path differences from the four-dimensional data during reconstruction. This enables the acquisition of high-quality, large-scale images, facilitating a more comprehensive understanding and analysis of materials microstructure. Moreover, PDP has the potential to promote the widespread application of ptychographic tomography in the analysis of planar samples.

# Introduction:

In the theory and experiments of electron microscopy, it is typically assumed that the propagation direction of the electron beam is parallel to the normal direction of the sample surface. However, this assumption does not hold true in most experiments. Firstly, the surface of the thin sample in transmission electron microscopy (TEM) is usually tilted relative to the optical axis direction. Secondly, in atomic resolution experiments, to ensure that the crystallographic zone axis of the sample is as parallel as possible to the optical axis of the microscope[1-4], planar samples are often tilted within a limited angular range, generally within 30°. Thirdly, in atomic electron tomography (AET) experiments, the sample needs to be continuously tilted at larger angles, typically around ±70° [5-7]. All these factors cause the normal of the sample surface not to be parallel to the propagation direction of the electron beam, resulting in optical path differences (OPD, $D$) or defocus between different positions of the electron wave function, as shown in **Fig. 1**.

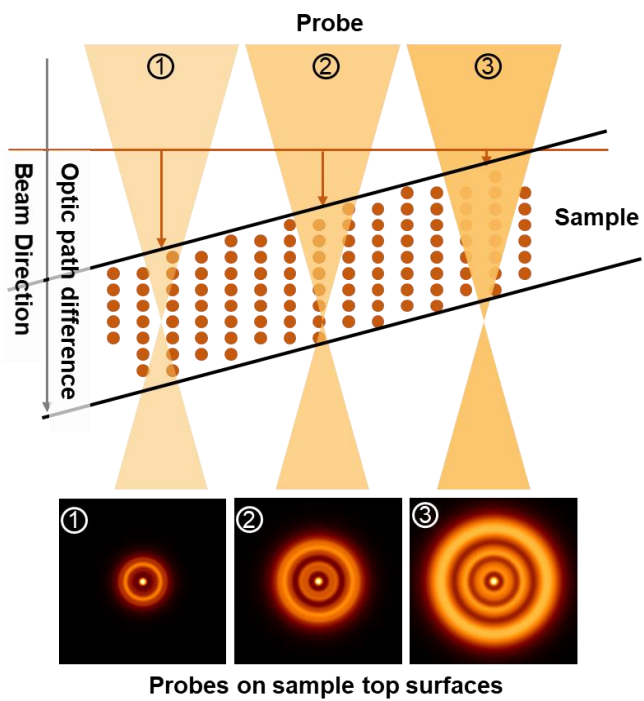

**Fig. 1. Diagram illustrating optical path differences in transmission electron microscopy experiments.** In transmission electron microscopy, to ensure that the crystallographic zone axis is parallel to the electron beam, there is often a deviation between the normal to the sample surface and the direction of the electron beam. This results in optical path differences when the electron

beam reaches the upper surface of the sample, leading to variations of electron beam at different positions.

Defocus is one of the critical experimental parameters in optical experiments and often requires careful selection [8]. Each experimental technique corresponds to a different optimal imaging defocus [8-12]. For example, in high-angle annular dark-field (HAADF) imaging, the best imaging results are achieved when the electron beam is focused on the sample surface [9,12]. However, when the sample plane is tilted at a large angle, due to the limited depth of field of the electron beam, clear images can only be obtained within a small range. In regions away from the optimal defocus position, the imaging quality deteriorates. This can have a significant negative impact on AET, which uses a series of tilted HAADF images, thereby limiting its application in planar samples.

Electron ptychography, as a computational imaging method, uses two-dimensional diffraction patterns at each scanning point in a two-dimensional sample plane (i.e., 4D data) [13,14]. Through algorithmic iterations, it can decouple the illumination beam and object phase information [15-17], thereby achieving high-precision, high-resolution, aberration-free imaging results. This technique has gained significant attention in recent years [4,18-23]. In the field of electron microscopy, Jiang et al. first utilized iterative electron ptychography to achieve a 39 pm information limit in 2D materials [20]. Subsequently, multi-state [24] and multi-slice [16] approaches were introduced into the electron ptychography algorithm, achieving a 22 pm information limit in bulk materials [21,22]. In 2022, Sha et al. introduced an adaptive propagator into iterative ptychography, overcoming the issue of sample zone axis misorientation that cannot be entirely avoided in experiments [4]. This enhancement improved the robustness of ptychography and has already found widespread applications in various fields, such as sub-nanometer resolution imaging of radiation-sensitive materials [10], three-dimensional characterization of dislocations [25], and direct detection of antiferromagnetic signals [26]. In 2024, Yang et al. took a novel approach by using localized orbital parameterization to describe the sample and electron beam, which significantly reduced the optimization parameters of the ptychography algorithm, achieving an ultimate resolution of 14 pm [23].

Although electron ptychographic algorithms excel in imaging performance [4,22,23], traditional ptychographic algorithms assume the electron beam at each scanning position is the same, ignoring changes in the optical path (i.e., Path fixed ptychography, PFP). This oversight has minimal impact on the reconstruction of small range of sample. However, in two crucial areas, neglecting optical path differences is unacceptable.

First, the advent of next-generation high-speed direct electron detectors [27,28] has significantly increased frame rates, making it possible to acquire large-scale 4D data. Second, ptychographic tomography is a vital direction for atomic-scale three-dimensional imaging, typically requiring large-angle tilting of the sample. In both cases, ignoring optical path differences leads to poor reconstruction quality and artifacts in the results.

To address these issues, we propose the path difference ptychography (PDP), which incorporates the optical path difference as optimization variables in the iterative process of ptychographic imaging. Experimental data confirm that even under conditions where the sample plane is tilted by 22.5°, PDP can produce high-quality, uniform large-area imaging results. PDP will play a crucial role in large-scale ptychographic imaging and ptychographic tomography.

## Results and discuss:

**Theory of PDP:**

In TEM experiments, there are three critical directions: the incident direction of the electron beam (i.e., the optical axis direction), the crystallographic zone axis direction of the sample, and the normal direction of the sample surface. Ideally, these three directions should be parallel to each other to ensure optimal imaging conditions. However, in practical situations, regardless of the sample preparation technique used, there is always some deviation between the normal direction of the sample surface and the zone axis. To align the crystallographic zone axis of the sample with the incident direction of the electron beam, it is often necessary to sacrifice the parallelism between the sample surface normal and the electron beam incident direction. This trade-off inevitably introduces OPD to electron beam at different positions of the sample (**Fig. 1**).

As for scanning transmission electron microscopy (STEM) imaging, the presence of OPD results in varying defocus amounts at different scan positions. This directly affects the state of the electron beam (**Fig. 1**). As the scan position changes, the shape and intensity of the electron beam undergo significant alterations, adversely impacting the final image quality. This not only reduces image clarity but can also introduce artifacts, hindering the accurate analysis of the sample's microstructure.

The core of the iterative ptychographic imaging algorithm lies in two interdependent processes [15-17]: forward propagation and backward optimization. The forward propagation process is responsible for simulating the diffraction data, while the backward optimization step iteratively refines a series of parameters based on the discrepancies between the experimental data and the simulated diffraction data produced in forward propagation step. These parameters include, but are not limited to, the phase distribution of the object, the illumination beam, and the scan positions [15,23,29]. The performance of the algorithm is closely related to the thoroughness with which the forward propagation process is described [4,29]. A more detailed and comprehensive forward propagation model can significantly enhance the stability and convergence speed of the algorithm, allowing for the extraction of deeper information from the data. For example, introducing an adaptive propagator during the forward propagation process [4] not only improves the robustness of the algorithm but also enables to acquire the information about the zone axis deviation.

To construct a more accurate forward propagation model, we incorporate the optical path difference ($D$) as an optimization variable into the forward propagation stage of the iterative ptychographic imaging technique (as illustrated in the flowchart in **Fig. 2**). Before calculating the interaction between the electron beam and the sample, we account for the influence of $D$, thus generating differentiated electron beams for different scan positions:

$$\boldsymbol{P}_{r^j} = \mathcal{F}^{-1}\{\mathcal{F}\{\boldsymbol{P}(r)\} \cdot \exp(2\pi i \boldsymbol{k} \cdot \boldsymbol{r}^j) \cdot \exp(-i\pi D_{r^j}\lambda|\boldsymbol{k}|^2)\}$$

Here, $j$ is the scan point index, $\boldsymbol{r}^j$ is the position of the $j$-th scan point, $\boldsymbol{P}(r)$ is the initial electron beam function independent of the scan position, $\mathcal{F}$ and $\mathcal{F}^{-1}$ represent the Fourier transform and inverse Fourier transform, respectively, $\boldsymbol{k}$ represents the spatial frequency, $D_{r^j}$ is the optical path difference at the scan position $\boldsymbol{r}^j$, and $\lambda$ is the wavelength of the electron beam. The last two terms represent the translation of the electron beam within the sample plane and the influence of the optical path difference at each scan position, respectively.

Subsequently, using the differentiated electron beam $\boldsymbol{P}_{r^j}$ and the object function $\boldsymbol{O}$, we calculate the simulated exit wave function according to the multislice method:

$$\varphi_{ext} = \mathbb{P}_{\Delta z}\{...\mathbb{P}_{\Delta z}\{\mathbb{P}_{\Delta z}\{\boldsymbol{P}_{r^j}\boldsymbol{O}_1(r)\}\boldsymbol{O}_2(r)\}\boldsymbol{O}_3(r)...\}\boldsymbol{O}_L(r)$$

Here, the Fresnel near-field diffraction factor $\mathcal{P}_{\Delta z}\{\bullet\}$ is given by [4]:

$$\mathcal{P}_{\Delta z}\{\bullet\} = \mathcal{F}^{-1}\{\mathcal{F}\{\bullet\}(k)\exp(-i\pi\Delta z(\lambda|k|^2 - 2k_x\tan\theta_x - 2k_y\tan\theta_y))\}$$

which describes the propagation between slices, where $\Delta z$ is the slice thickness, $\boldsymbol{\theta} = (\theta_x, \theta_y)$ is the misorientation between the crystallographic zone axis and the beam incident direction. $\boldsymbol{O}_l(r)$ represents the object function of the $l$-th slice. For a single-slice ptychographic imaging method, $L = 1$.

Next, based on the loss function $\mathcal{L}$ derived from the difference between the simulated diffraction data and the experimental data [4,30,31], we can optimize the gradients of the object function $\boldsymbol{O}(r)$, the electron beam function $\boldsymbol{P}(r)$, and the optical path difference $D$. This process ultimately can decouple the optical path difference from the diffraction data, resulting in higher quality reconstruction results.

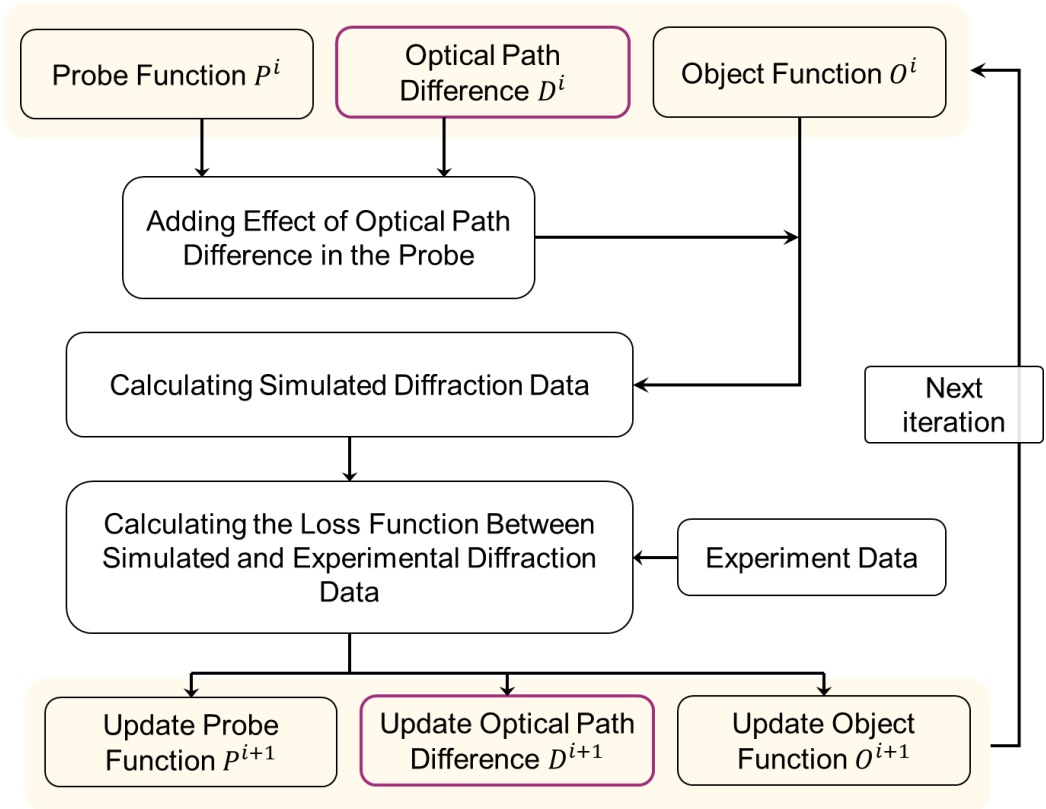

**Fig. 2. Flowchart of each iteration process in PDP.** Unlike PFP, PDP incorporates the impact of optical path difference and optimizes it during the backward iteration process.

**Large area reconstruction results:**

In the field of electron microscopy, most direct imaging and computational imaging techniques are based on the assumption of a constant and unchanging illumination beam. This makes the OPD a major limiting factor affecting image quality. However, by using the PDP method, it is possible to effectively eliminate the interference caused by OPD between different scan positions. This significantly enhances the accuracy of the imaging results.

To visually demonstrate the advantages of the PDP technique, we followed the sample preparation process described in the supplementary materials to create a planar sample with a surface normal direction deviating by 22.5° from the crystallographic zone axis. Using STEM mode, we performed comprehensive characterization of this sample, collecting a HAADF image covering a 36-nanometer range (**Fig. 3**a), along with the corresponding 4D diffraction dataset. Subsequently, we applied both the PDP and PFP to reconstruct the 4D data. The comparison of the results is presented in **Fig. 3**b and **Fig. 3**c, respectively. Detailed experimental acquisition conditions and reconstruction parameter settings can be found in the supplementary materials.

For the HAADF image shown in **Fig. 3**a, the electron beam is focused only on the central region, resulting in a clear image in the center but blurry areas towards the edges due to defocusing. This phenomenon highlights the difficulty of achieving high-quality large-area imaging through HAADF when there is OPD caused by the sample plane's tilt. Both **Fig. 3**b and **Fig. 3**c present a comparison of the reconstructed results for the sample region shown in **Fig. 3**a, using the PFP and PDP, respectively.

PFP fails to effectively account for the optical path difference, resulting in a reconstructed electron beam function (Fig. S5b) that exhibits noticeable mess, ultimately compromising imaging quality. The corresponding object function (**Fig. 3**b) displays a low contrast and a poor signal-to-noise ratio. In contrast, PDP significantly enhances the quality of both the electron beam (Fig. S5a) and the object function (**Fig. 3**c). Moreover, PDP accurately reconstructs parameters related to the tilt of surface normal (Fig. S2), and the distribution of the OPD at different sample positions (**Fig. 3**d). This indicates that PDP not only improves imaging quality but also provides additional information about the sample's geometric parameters, which is beneficial for understanding samples' structures.

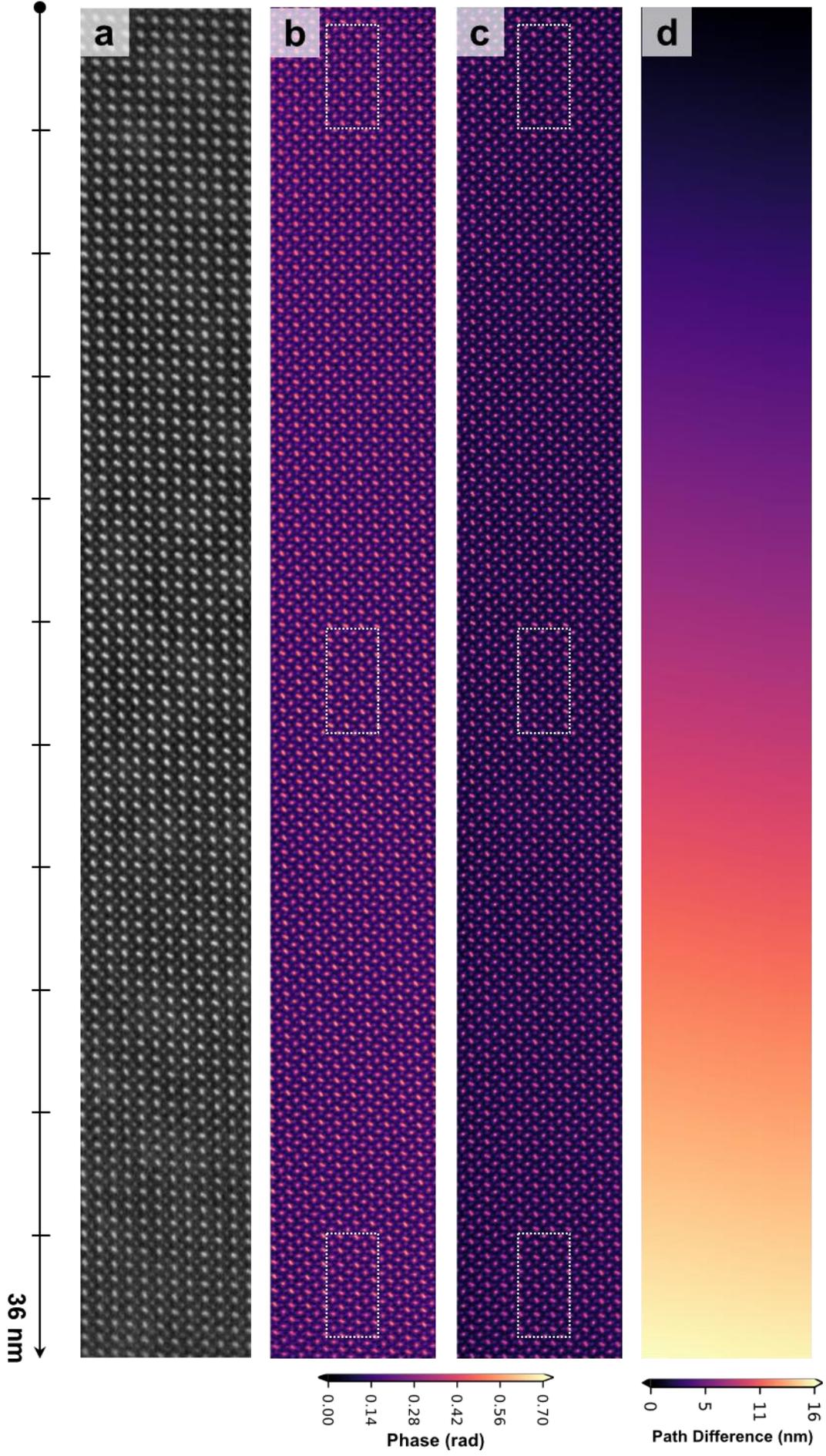

**Fig. 3. Display of PDP large-area reconstruction results.** (a) HAADF image, (b) PFP result, (c) PDP reconstruction result, (d) optical path difference for the region obtained through PDP. (a), (b), and (c) correspond to the same sampling area, with (b) and (c) representing reconstruction results from the same dataset.

**Quality analysis of reconstruction results:**

The left, middle, and right regions in **Fig. 3**b and **Fig. 3**c are enlarged and shown in **Fig. 4**. It is evident that, compared to the PFP method, the PDP technique exhibits superior reconstruction quality and demonstrates higher overall uniformity. The impact of OPD results in variable image quality with the PFP method, where the signal-to-noise ratio and imaging clarity in the both sides significantly worse than that of the central region. In contrast, the PDP results maintain high quality, a strong signal-to-noise ratio, and excellent uniformity across the entire field of view.

It is noteworthy that the comparison of diffractograms in **Fig. 4**a to **Fig. 4**f indicates that the information limit achievable with the PDP far surpasses that of the PFP, showing nearly a two-fold improvement, from 0.6 Å to 0.3 Å. **Fig. 4**g to **Fig. 4**l depict the phase intensity curves for Dy-Dy, O-Sc-O, and O atoms, respectively, clearly demonstrating that the atomic peaks reconstructed using the PDP have narrower full width at half maximum. This signifies higher reconstruction accuracy and image quality.

For the three curves that should theoretically exhibit symmetrical distribution, the reconstruction results from the PFP show noticeable artifacts, whereas the PDP effectively avoids such issues, providing more accurate and reliable data. This series of comparative analyses clearly demonstrates the significant advantage of the PDP in enhancing imaging quality.

Large-scale high-quality imaging techniques play a crucial role in materials science research [32,33], especially in revealing low-frequency signal characteristics of materials. However, when dealing with extensive sample areas, the significant increase in optical path differences presents a considerable challenge, potentially leading to severe degradation in image quality, particularly in imaging complex structures. In this context, the advent of the PDP provides a powerful tool to address this issue. By effectively decoupling optical path differences, PDP can significantly enhance the quality and uniformity of large area reconstructed images. Moreover, current AET is limited to planar samples and is only applicable to particulate or needle-like samples [34,35]. We anticipate that the PDP technique, being applicable with sample high angle tilt, will strongly support the broader application of tomography in planar samples.

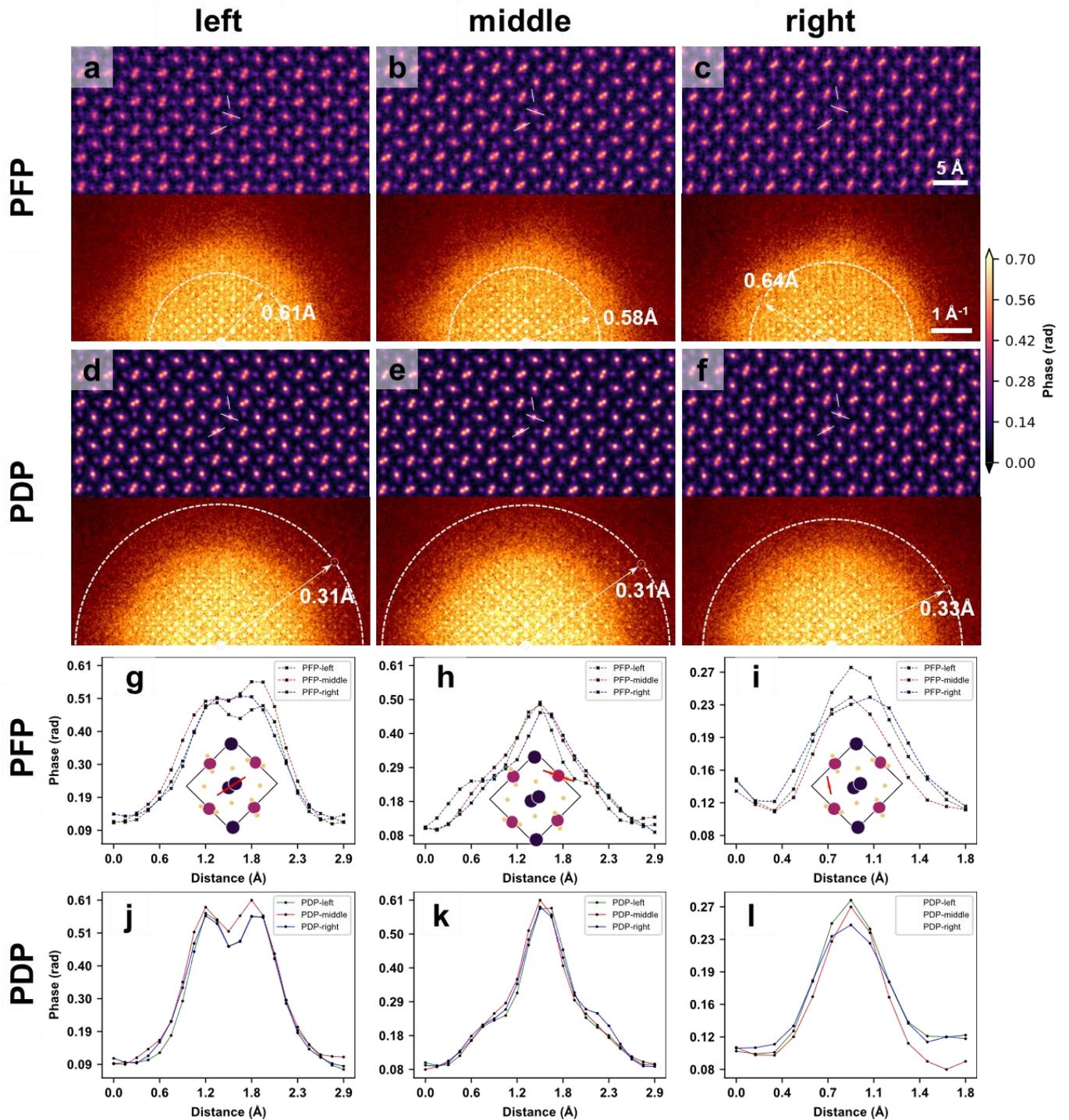

**Fig. 4. Comparison of PDP and PFP reconstruction results.** (a-f) show results from the boxed regions in **Fig. 3** for both PFP and PDP, with the corresponding diffraction patterns displayed in the lower part of each image. Compared to the PFP results, the PDP reconstructions are more uniform and achieve twice the information limit of the PFP results. (g-l) depict intensity curves of the white short lines shown in (a-f), with (g and j) representing Dy-Dy, (h and k) O-Sc-O, and (i and l) O.

**Revealing the surface morphology of the sample:**

PDP not only achieves higher precision and a broader field-of-view in reconstructed images but also reveals critical shape parameters of the sample's surface, providing valuable insights into its physical properties and enabling a more precise analysis of its structural features.

**Fig. 3**d and Fig. S2 demonstrate how PDP reveals surface shape information resulting from the tilt of surface normal. To further validate the ability of PDP to reconstruct intrinsic surface features of samples, we selected $PbZrO_3$ material with voids as a case study (crystal structure referenced in Fig. S4). By performing 4D data acquisition of this region and applying PDP reconstruction, we successfully revealed the pit structures located on the sample's upper surface (as shown in **Fig. 5**b). Importantly, after decoupling the optical path differences caused by the sample's intrinsic surface shape, we obtained high-quality image (**Fig. 5**a), which lays a solid foundation for subsequent structural analysis and highlights the unique value of PDP in material characterization.

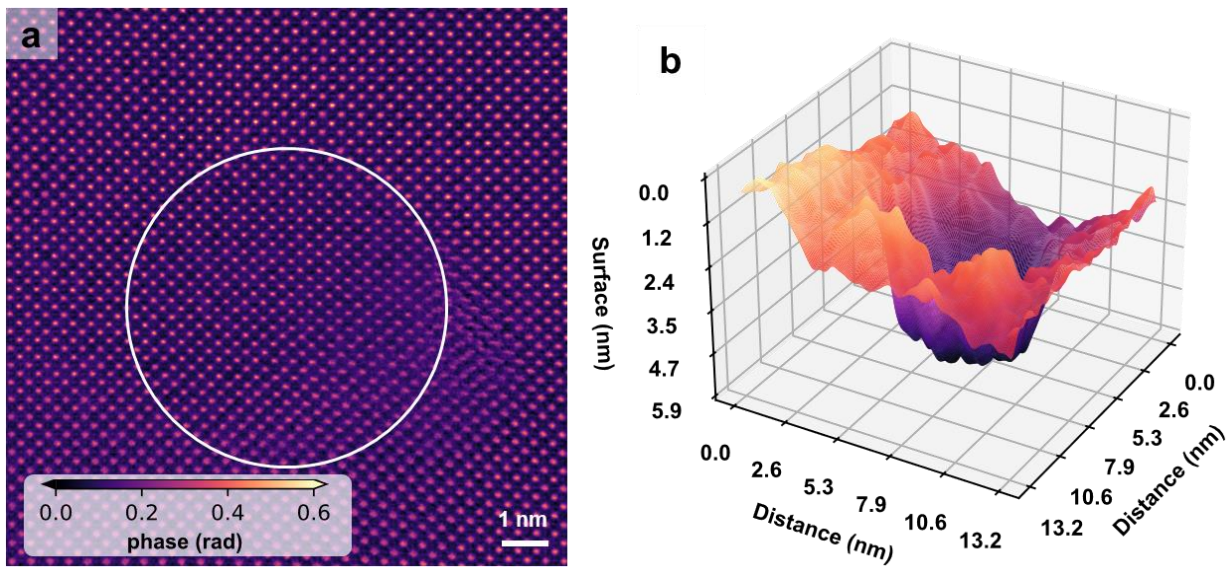

**Fig. 5. Surface information obtained using PDP.** For a sample with voids, PDP can reconstruct a high-quality object function (a) and the sample's shape (b). The white circles in (a) highlight the areas where the voids occur, resulting in an optical path difference of 6 nm.

## Conclusions:

Optical path difference is a common factor in electron microscopy imaging and poses a significant barrier to obtaining clear imaging results over large areas. To address this challenge, we incorporated optical path difference parameters into the reconstruction process of ptychography. By optimizing this variable, we developed the path difference ptychography. Its core advantage lies in its ability to effectively eliminate the negative impact of optical path differences, thereby achieving uniformly high quality and large-scale imaging. This provides a powerful tool for precise structural analysis of large-scale and high-throughput.

Furthermore, the PDP technique is not limited to enhancing image quality, it also provides additional geometric information, such as the shape of the sample and the surface morphologies. It is particularly noteworthy that tomography has so far been primarily applied to the analysis of nanoparticles or nano-needles. However, most TEM samples are planar shaped, tilting samples would introduce different path difference for each tilt angle, resulting in lowered quality in the ptychographic phase images to a different extent, posing a big challenge to the application of ptychographic tomography. Incorporating optical path difference in ptychography helps to remove the obstacle and promotes widespread applications of ptychographic tomography in materials science.

## Acknowledgments:

In this work we used the resources of the Physical Sciences Center and Center of High-Performance Computing, Tsinghua University.

## Funding:

This work was supported by the National Natural Science Foundation of China (52388201 and 51525102) and the Postdoctoral Fellowship Program (Grade B) of China Postdoctoral Science Foundation (GZB20240329).

## Author contributions:

## Competing interests:

The authors declare no competing financial interests.

## Supplementary materials

**Sample preparation:**

In this experiment, a cross-sectional sample of DyScO$_3$ was prepared with a 22.5° deviation between the sample's normal direction and the zone axes using focused ion beam (FIB) technology, as illustrated in Fig. S1. The DyScO$_3$ single crystal was procured from Hefei Kejing Materials Technology Co., Ltd. Similarly, the PbZrO$_3$ sample shown in **Fig. 5** was also prepared using FIB. During the sample preparation, the specimens were thinned to approximately 20 nm using an accelerating voltage of 30 kV and a decreasing current ranging from 240 to 50 pA. This process was followed by fine polishing with an accelerating voltage of 5 kV and a current of 20 pA. The final sample thickness, as determined by ptychography, was 15 nm.

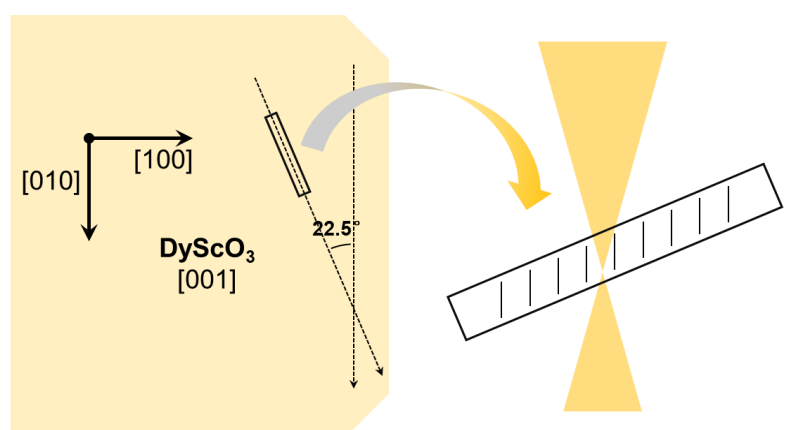

Fig. S1. Schematic diagram of a sample with a deviation between the surface normal and the crystallographic zone axis. The cross-sectional sample is extracted after rotating clockwise 22.5 degrees relative to the [001] direction.

**STEM experiments and ptychographic reconstructions:**

The HAADF and four-dimensional scanning diffraction (4D) datasets were acquired using a probe aberration-corrected FEI Titan Cubed Themis G2 operated at 300 kV. The convergence semi-angle was 25 mrad, and the HAADF collection semi-angle was 49-200 mrad. 4D-STEM data were collected using a pixel-array detector (EMPAD), with each diffraction pattern comprising 128 × 128 pixels. A nominal camera length of 460 mm yielded a reciprocal space pixel size of 0.034 Å$^{-1}$. An underfocus value of 20 nm, determined from the optimal HAADF image of the middle area, was selected. A scan step size of 0.74 Å was employed. The beam current was 15 pA, and the dwell time was 0.5 ms. For ptychographic reconstruction, diffraction patterns were zero-padded to 200 × 200 pixels, resulting in a real-space pixel size of 0.14 Å.

**Supplementary Figures:**

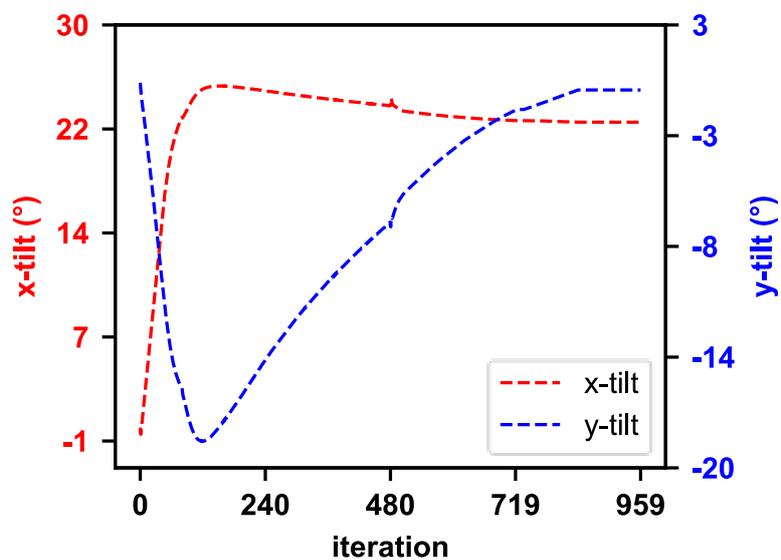

Fig. S2. Angle deviation between the sample normal direction and the electron beam as updated obtained from PDP.

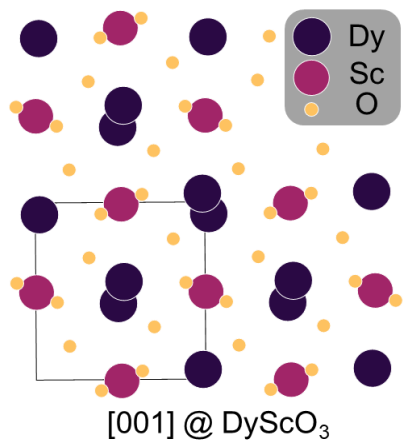

Fig. S3. Structure of $DyScO_3$.

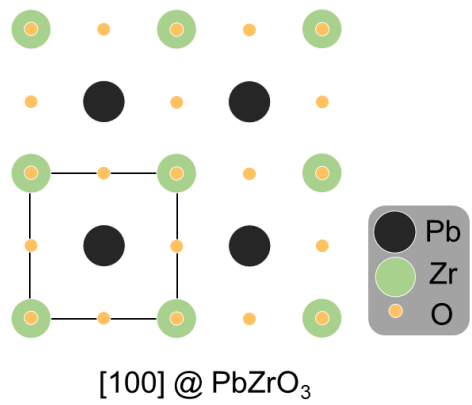

[100] @ PbZrO$_3$

Fig. S4. Structure of PbZrO$_3$.

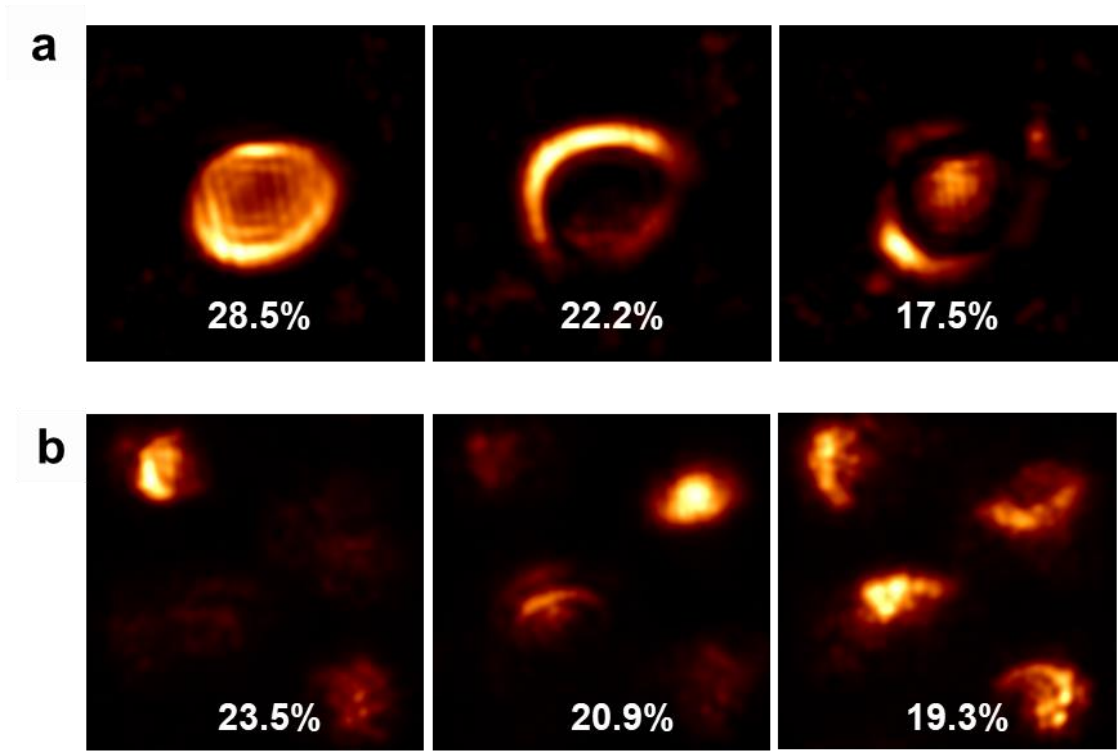

Fig. S5. Mixed electron beam states for PDP and PFP. (a) and (b) show the first three modes of the electron beam for PDP and PFP, respectively, with the corresponding mode contributions labeled.